\documentclass[twocolumn,showpacs,preprintnumbers,amsmath,amssymb,aps,superscriptaddress]{revtex4}

\usepackage{graphicx}
\usepackage{dcolumn}
\usepackage{bm}
\newcommand{\bq}{\begin{eqnarray}}
\newcommand{\eq}{\end{eqnarray}}

\begin{document}

\title{Critical fluctuations in a soliton formation of attractive Bose-Einstein condensates}

\author{Rina Kanamoto}
\affiliation{Department of Physics, Osaka City University, Osaka 558-8585, Japan}

\author{Hiroki Saito}
\affiliation{Department of Physics, Tokyo Institute of Technology, Tokyo 152-8551, Japan}

\author{Masahito Ueda}
\affiliation{Department of Physics, Tokyo Institute of Technology, Tokyo 152-8551, Japan}
\affiliation{ERATO, Japan Science and Technology Corporation (JST), Saitama 332-0012, Japan}

\date{\today}

\begin{abstract}
We employ mean-field, Bogoliubov and many-body theories to study 
critical fluctuations in the position and momentum 
of a Bose-Einstein condensate whose translation symmetry 
is spontaneously broken due to attractive interactions. 
In a homogeneous system, the many-body ground state of 
the symmetry-preserving Hamiltonian is very fragile 
against superposition of low-lying states, while 
the mean-field theory predicts a stable bright soliton which 
spontaneously breaks translation symmetry. 
We show that weak symmetry-breaking perturbations 
cause the translation-symmetric many-body ground state 
to cross over to a many-body bright soliton. 
We argue that the center-of-mass fluctuations in the soliton state 
arise primarily from the depletion of the condensate to translation modes. 
We develop an extended mean-field theory to analytically reproduce these 
results obtained by the exact diagonalization method. 
\end{abstract}
\maketitle

\section{Introduction}\label{sec1}

Ultracold matter waves offer the possibility of observing 
quantum-mechanical fluctuations associated 
with low-dimensional many-body effects 
and phase diffusions~\cite{Aspect,Arlt,Orzel,EW,LY}. 
In Bose-Einstein condensates (BECs) of dilute gases, 
various techniques of controlling experimental parameters 
enable us to study a rich variety of phase transitions. 
A standard way to study a ground-state phase of the condensate is 
to introduce an order parameter with spontaneously broken symmetry. 
On the other hand, quantum fluctuations play a crucial role in the emergence of 
the order parameter, and the related issues have been widely studied in 
the systems of BECs~\cite{LPB,KU,HY,spin1,AL,spin-half,SB-Vor,SB-Rot}. 
Furthermore, since atomic condensates are on the mesoscopic scale, 
we may expect interesting finite-size effects which are absent 
in the thermodynamic limit. 

The exact many-body ground state of a one-dimensional (1D) system 
with translation symmetry does not exhibit the off-diagonal long-range 
order in thermodynamic limit. 
The ground state has the translation symmetry of the original Hamiltonian, 
but is extremely fragile against localization to a bright soliton 
when the interaction is attractive. 
By investigating the many-body energy spectrum of this system, we have 
found that there exists a large number of low-lying quasidegenerate 
many-body eigenstates, and the localization to a soliton is due to 
the superposition of these states~\cite{KSU}. 
In the thermodynamic limit, the energy gap between the ground and low-lying 
states vanishes, and therefore the localization occurs no matter how small 
the symmetry-breaking perturbation is. 
Such a localized soliton~\cite{MFsoliton} can be well described by the Gross-Pitaevskii (GP) equation. 
However, in a mesoscopic system, an energy gap between the ground 
and low-lying excited states remains finite, and therefore we can expect a crossover 
between the state with the translation symmetry and the soliton state 
as a function of the strength of interaction. 
In such a crossover regime, quantum fluctuations are expected to be large and 
the localization process is nontrivial. 

The aim of this paper is to examine the critical fluctuations 
in a soliton formation of an attractive BEC. 
We consider two cases for the breaking of translation symmetry, 
that is, symmetry-breaking potentials and quantum measurements. 
We tackle this problem using the numerical diagonalization 
of the Hamiltonian and an extended mean-field theory. 
In particular, we explicitly calculate the uncertainty relation between 
the center-of-mass position and momentum, and show that 
the inclusion of center-of-mass fluctuations 
of the soliton is crucial for understanding the nature of 
the many-body ground state. 
These fluctuations emerge as a depletion of the condensate, 
and decrease with increasing the magnitude of the symmetry-breaking 
potential or the number of the measurements. 

This paper is organized as follows. 
In Sec.~\ref{QPT}, we introduce a model of a 1D attractive Bose system 
with the periodic boundary condition, and briefly review 
the many-body energy spectrum and ground-state properties of the uniform system. 
In Sec.~\ref{MBSBP}, we introduce as symmetry-breaking perturbations 
a symmetry-breaking potential, a double-well potential, and 
quantum measurements, and discuss critical fluctuations in the position and 
momentum of symmetry-broken states. 
In Sec.~\ref{MFBog}, we construct the Bogoliubov ground state 
that incorporates the effect of center-of-mass fluctuations 
due to the presence of the infinitesimal potential. 
We also propose an extended mean-field theory to describe the effect of 
quantum measurements on the uncertain relation between the 
position and the momentum. In Sec.~\ref{Sec5}, we summarize 
the main results of this paper. 


\section{Quantum Phase Transition in An Attractive BEC}\label{QPT}

\subsection{Many-body eigenstates}\label{QPT1}

\begin{figure}
\includegraphics[scale=0.45]{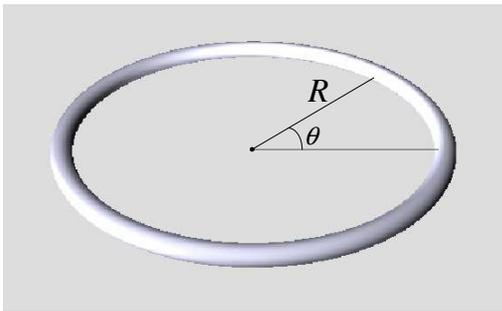}
\caption{
(Color online) Schematic illustration of a quasi-one-dimensional torus. 
}
\label{ring}
\end{figure}

We consider a system of weakly interacting bosons on 
a quasi-1D torus with radius $R$ as 
schematically illustrated in Fig.~\ref{ring}. 
Throughout this paper the length, angular momentum, and energy are measured in units of 
$R$, $\hbar$, and $\hbar^2/(2\pi R)$, respectively. 
The Hamiltonian of the system is given by
\begin{eqnarray}\label{Hamiltonian}
\hat{H}=\int_0^{2\pi}d\theta
\left[-\hat{\psi}^{\dagger}(\theta)\frac{\partial^2}{\partial\theta^2}\hat{\psi}(\theta)
-\frac{\pi g}{2}\hat{\psi}^{\dagger 2}(\theta)\hat{\psi}^2(\theta)\right],
\end{eqnarray}
where $\theta$ is the azimuthal angle, 
$\hat{\psi}$ is the boson field operator obeying 
the periodic boundary condition $\hat{\psi}(\theta)=\hat{\psi}(\theta+2\pi)$, 
and the interaction is assumed to be attractive $(g>0)$. 
Although the Hamiltonian~(\ref{Hamiltonian}) is exactly solvable and 
the exact solution is analytically obtained by the Bethe ansatz~\cite{C}, 
in principle, the solution is intractable for a large number of atoms. 
On the other hand, the configuration interaction method can evaluate 
the ground-state energy rather accurately~\cite{ASC}. 
Experimentally the quasi-1D ring-shaped BEC has recently been realized in 
a circular magnetic trap~\cite{ring}. 

We employ a numerical diagonalization method to 
investigate many-body eigenstates of the Hamiltonian~(\ref{Hamiltonian}) 
by restricting the Hilbert space to that spanned by the angular-momentum $l=0,\pm 1$ states. 
It is possible to extend the calculation to including the $l=\pm 2,\pm 3$ states. 
However, all results obtained for $l=0,\pm 1$ remain qualitatively unchanged, and 
we thus restrict ourselves to $l=0, \pm 1$ unless otherwise stated (see Sec.~\ref{MB_pot2}). 
The Fock-state bases are thus written as $|n_{-1},n_0,n_1\rangle$, where 
$n_l$ is the number of atoms with angular momentum $l$. 
The number of atoms and the total angular momentum are given by 
$\sum_l n_l=N$, and $\sum_l l n_l ={\cal L}$, respectively. 
The field operator is thus expanded as
\bq
\hat{\psi}(\theta)=\frac{1}{\sqrt{2\pi}}(\hat{c}_0+\hat{c}_1e^{i\theta}+\hat{c}_{-1}e^{-i\theta}),
\eq
where $\hat{c}_l$ is the annihilation operator of a boson with angular momentum $l$. 

\begin{figure}[b]
\includegraphics[scale=0.45]{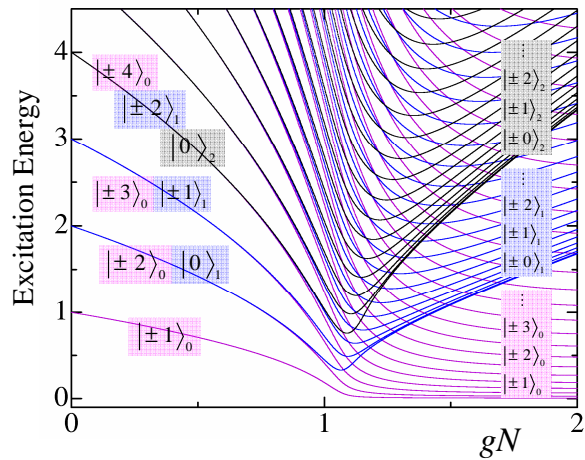}
\caption{
(Color online) Many-body excitation spectrum $E_{\sigma,{\cal L}}-E_{0,0}$ of 
the Hamiltonian $\hat{H}$ with $N=200$, 
where ${\cal L}$ is the angular-momentum index and $\sigma$ 
is the index of the bands that appears for $gN \gtrsim 1$. 
We plot only low-lying states with indices $\sigma=0,\ 1$, and $2$. 
The state $|{\cal L}\rangle_{\sigma}$ denotes the many-body 
eigenstate whose total angular momentum is ${\cal L}$ and band index is $\sigma$.
}
\label{fig_sec2}
\end{figure}

Figure~\ref{fig_sec2} shows the low-lying many-body spectrum 
of the Hamiltonian $\hat{H}$, in which extensive rearrangement of the 
distribution of the eigenstates is seen to occur at $gN\simeq 1$ due to the quantum phase transition. 
We see that in $gN\gtrsim 1$ the spectrum is classified by 
two indices: that is, the band index $\sigma$ and the angular-momentum index ${\cal L}$ 
characterize each level as 
\bq
\hat{H}|{\cal L}\rangle_{\sigma}
=E_{\sigma,{\cal L}}|{\cal L}\rangle_{\sigma},
\eq
where
\bq
|{\cal L}\rangle_{\sigma}=\sum_n A_n |n,N-2n-{\cal L},n+{\cal L}\rangle\label{stateL}. 
\eq
The energy levels $E_{\sigma,{\cal L}}$ 
in band $\sigma$ are distributed according to 
$E_{\sigma,0}\lesssim E_{\sigma,\pm 1}\lesssim E_{\sigma,\pm2},...,$ . 
For $0\le gN \lesssim 1$, on the other hand, 
some $E_{\sigma,{\cal L}}$'s are almost degenerate. 
To understand the nature of the rearrangement at $gN \simeq 1$, we first study the eigenstates 
in the absence of interaction $gN=0$, where the eigenstates can be 
described by Fock states $|n_{-1},n_0,n_1\rangle$. 
We define the $J$th state ($J=0,1,2,\dots,)$ as the one in which $J$ atoms 
are excited. For examples, the $J=0$ state corresponds to the noninteracting 
ground state given by 
\bq 
J=0 : |0\rangle_{\sigma=0}=|0,N,0\rangle, 
\eq
and the excited states are given as 
\bq
J=1 &: & |-1\rangle_{\sigma=0}=|1,N-1,0\rangle,\nonumber\\
J=1 &: & |1\rangle_{\sigma=0}=|0,N-1,1\rangle,\nonumber\\
J=2 &: & |-2\rangle_{\sigma=0}=|2,N-2,0\rangle,\nonumber\\
J=2 &: & |0\rangle_{\sigma=1}=|1,N-2,1\rangle,\nonumber\\
J=2 &: & |2\rangle_{\sigma=0}=|0,N-2,2\rangle,\nonumber\\
J=3 &: & |-3\rangle_{\sigma=0}=|3,N-3,0\rangle,\nonumber\\
\ &\vdots&.
\eq
The $J$th state is hence $(J+1)$-fold degenerate. 
The index $\sigma$, which is the band index in $gN \gtrsim 1$, corresponds to 
the number of $l=\pm 1$ pairs in the Fock state for each $J$th state at $gN=0$, i.e., 
$\sigma \equiv (J-|{\cal L}|)/2=0,1,2,\dots,$ . 
For $0<gN\lesssim 1$, the levels are characterized by $J$ alone, 
since the degeneracy with respect to $\sigma$ at $gN=0$ is maintained. 

The excitations in $0\le gN \lesssim 1$ have 
substantial energy gaps until the critical point $gN\simeq 1$ is reached. 
On the other hand, for $gN\gtrsim 1$, the energy difference between the ground and the 
first excited states scales as $1/N$. 
We thus expect that the ground state for $gN \gtrsim 1$ is vulnerable for large $N$, 
and perturbations of the order $1/N$ can cause drastic reconstruction of 
the ground state. 
The emergence of these quasidegeneracies in the energy spectrum may be regarded as 
a precursor of symmetry breaking, and interesting effects such as 
the enhancement of the condensate fraction~\cite{KSU} can be expected 
as a consequence of symmetry breaking. 


\subsection{Ground-state properties}\label{QPT2}

\begin{figure}[t]
\includegraphics[scale=0.5]{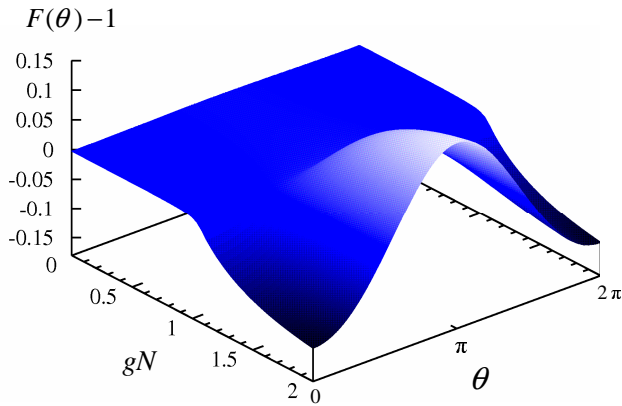}
\caption{
(Color online) Fano factor $F(\theta)$ in Eq.~(\ref{fano}) of the many-body ground-state 
wave function as functions of the strength of the interaction $gN$ and azimuthal angle $\theta$ 
with $N=200$. 
}
\label{fig_fano}
\end{figure}

We study properties of the ground state $|{\cal L}=0 \rangle_{\sigma=0}$ 
as a function of the strength of interaction $gN$. 
According to the standard definition of the BEC~\cite{PO}, 
the condensate fraction and 
the condensate wave function for a many-body state $|\Psi\rangle$ are given in terms of 
the maximum eigenvalue $\lambda_{\rm M}$ 
and the corresponding eigenfunction of the reduced single-particle density matrix 
$\rho(\theta',\theta)= \langle \Psi|\hat{\psi}^{\dagger}(\theta')\hat{\psi}(\theta)|\Psi \rangle/N$. 
If the maximum eigenvalue $\lambda_{\rm M}$ is on the order of unity and the other ones are on 
the order of $1/N$, there exists a usual single condensate described by the corresponding 
eigenfunction. 
For the case of $gN \gtrsim 1$, however, the ground state does not fall into 
this category of BEC. 
By calculating the eigenvalue of $\rho$ for the ground state 
as a function of $gN$, 
we find that $\lambda_{\rm M}$ is on the order of 1 for $gN \lesssim 1$, 
but for $gN \gtrsim 1$ there appears more than one eigenvalue on the order of unity. 
The ground state is thus a conventional single condensate for $gN \lesssim 1$, and 
fragmented~\cite{NS} for $gN \gtrsim 1$. 

Since we consider the Hamiltonian with translation symmetry, 
the expectation value of the number density of particles $n(\theta)=N\rho(\theta,\theta)$ 
of the ground-state wave function is constant for $0\le \theta < 2\pi$.
In contrast, the two-body correlation function,
\bq
g^{(2)}(\theta,\theta')
=\frac{\langle\hat{\psi}^{\dagger}(\theta)\hat{\psi}^{\dagger}(\theta')\hat{\psi}(\theta')\hat{\psi}(\theta)\rangle}
{\langle\hat{\psi}^{\dagger}(\theta)\hat{\psi}(\theta)\rangle\langle\hat{\psi}^{\dagger}(\theta')\hat{\psi}(\theta')\rangle}, 
\eq
is found to deviate greatly from unity for $gN\gtrsim 1$, 
while it is almost unity for $gN\lesssim 1$~\cite{2003}. 
We show in Fig.~\ref{fig_fano} the Fano factor of the ground state defined by 
\bq\label{fano}
F(\theta)&\equiv&\frac{\langle n^2(\theta)\rangle-\langle n(\theta)\rangle^2}{n(\theta)}\nonumber\\
&=&1+n(\theta)\left[g^{(2)}(\theta,\theta)-1\right].
\eq
The large deviation of $F(\theta)$ from unity for $gN \gtrsim 1$ indicates 
that the number density of particles has quantum fluctuations, which may also be regarded 
as a precursor of formation of the broken-symmetry state. 


\section{Symmetry-Breaking Perturbations}\label{MBSBP}

In this section, we investigate on how the many-body ground state responds to 
infinitesimal symmetry-breaking potentials or quantum measurements. 

\subsection{Infinitesimal potential}\label{MB_pot1}


\subsubsection{Superposition of low-lying states}

We employ the exact diagonalization method to obtain the ground state of the Hamiltonian 
\bq\label{hamK}
\hat{K}_1=\hat{H}+\varepsilon_1\hat{V}_1,
\eq
where 
$\hat{V}_1=\int_0^{2\pi}d\theta \hat{\psi}^{\dagger}(\theta)\cos\theta\hat{\psi}(\theta)$ 
is a symmetry-breaking potential. 
Since we discuss here only low-lying modes ($\sigma=0, {\cal L}\ne 0$) 
close to the ground state ($\sigma=0, {\cal L}=0$), 
we simplify the notations of the eigenstates by omitting the index $\sigma$. 

We consider the energy change in the many-body ground state 
as a function of the magnitude $\varepsilon_1$ of the potential. 
Because of the degeneracy $E_{\cal L}=E_{-{\cal L}}$ with 
respect to the angular momentum, 
we assume that the broken-symmetry ground state 
$|\Psi^{\rm (\varepsilon_1)}_{\theta}\rangle$ of the Hamiltonian $\hat{K}_1$ 
is described by 
\bq\label{MBLoc}
|\Psi^{(\varepsilon_1)}_{\theta}\rangle= e^{-i{\hat L}\theta}
\left[\beta_0|0\rangle+\sum_{{\cal L}>0}
\beta_{\cal L}\left( |{\cal L}\rangle + |-{\cal L}\rangle\right)\right],
\eq
where the coefficients $\beta_{\cal L}$ 
satisfy the normalization condition, $\sum_{\cal L}|\beta_{\cal L}|^2=1$. 

\begin{figure}
\includegraphics[scale=0.43]{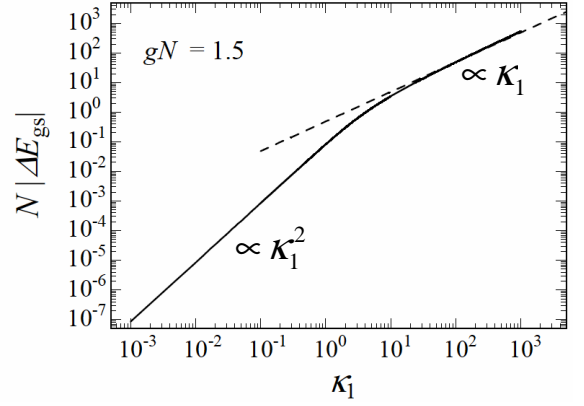}
\caption{
The amount of decrease in the ground-state energy given by Eq.~(\ref{Ene}) in the presence of 
a symmetry-breaking potential 
for $gN=1.5$, where $\kappa_1\equiv \varepsilon_1N^2$. 
The superimposed dashed line shows the result of the mean-field calculation given in Eq.~(\ref{GPene_dif}).
}
\label{fig_pot1_ene}
\end{figure}

\begin{figure}[b]
\includegraphics[scale=0.45]{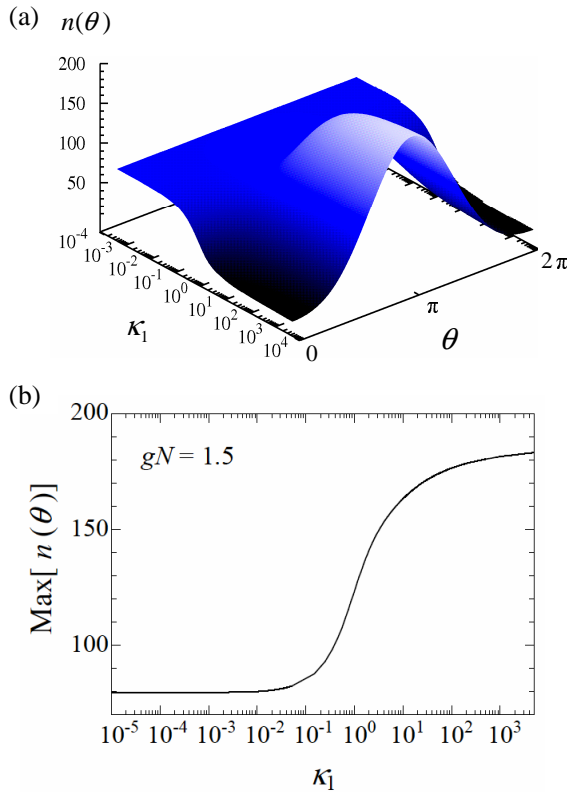}
\caption{
(Color online) (a) Density profile $n(\theta)$ of the many-body wave function and 
(b) the peak density as a function of $\kappa_1$. 
}
\label{fig_peak}
\end{figure}

In Fig.~\ref{fig_pot1_ene}, we show the energy change in the ground states
\bq\label{Ene}
\Delta E _{\rm gs}&\equiv&
\langle \Psi^{(\varepsilon_1)}_{\theta}|\hat{K}_1|\Psi^{(\varepsilon_1)}_\theta\rangle
-\langle 0|\hat{H}|0 \rangle\quad (<0),
\eq
multiplied by $N$, as a function of
\bq
\kappa_1\equiv \varepsilon_1 N^2.
\eq
The slope of the $\kappa_1$ dependence of $\Delta E_{\rm gs}$ changes near 
$\kappa_1\simeq 1$. 
The $\kappa_1$ dependence of $\Delta E_{\rm gs}$ for 
$0\le \kappa_1 \lesssim 1$ can be estimated by 
perturbation theory. The first-order correction to the ground-state energy 
is zero, and the second-order correction is given by 
\bq
E^{(2)}=\sum_{{\cal L}\ne 0}\frac{|\langle {\cal L}|\varepsilon_1 \hat{V}_1|0\rangle|^2}{E_0-E_{\cal L}},
\eq
where the denominator and the numerator are on the order of $N^{-1}$ and $(\varepsilon_1 N)^2$, respectively. 
The energy difference $\Delta E_{\rm gs}$ therefore depends on $-\varepsilon_1^2 N^3=-\kappa_1^2/N$. 
At the transition point $\kappa_1 \simeq 1$, 
the shift in the ground-state energy due to the symmetry-breaking 
perturbation is $\Delta |E_{\rm gs}|\simeq 1/N$, 
which is on the same order of magnitude as the energy gap. 
As $\kappa_1$ exceeds unity, the energy scale of the potential exceeds the energy gap $E_{\pm 1}-E_0$ 
and the matter wave begins to localize by superposing the ground 
and low-lying excited states as shown in Fig.~\ref{fig_peak}(a). 
The perturbation theory, therefore breaks down for $\kappa_1 \gtrsim 1$. 
The $\kappa_1$ dependence in this regime can be well described 
by the mean-field theory as shown in Sec.~\ref{GPBana}. 
In the absence of the symmetry-breaking potential, 
the formation of the broken-symmetry state $|\Psi^{(\varepsilon_1)}_{\theta}\rangle$ 
costs energy by an amount of 
\bq\label{Esup}
E_{\rm sup}&=&\langle\Psi^{(\varepsilon_1)}_{\theta}|\hat{H}|\Psi^{(\varepsilon_1)}_{\theta}\rangle\!-\!\langle 0|\hat{H}|0 \rangle\nonumber\\
&=&\sum_{\cal L}(E_{\cal L}-E_0)|\beta_{\cal L}|^2\quad (>0), 
\eq
which gives an energy increase associated with the superposition of the low-lying state. 

The symmetry breaking must be associated with 
a Nambu-Goldstone mode, 
which plays the role of restoring the symmetry. 
In the present context, the Nambu-Goldstone mode should be the 
translation zero mode $|\Phi^{(\varepsilon_1)}_{\theta}\rangle$ 
of the localized state $|\Psi^{(\varepsilon_1)}_{\theta}\rangle$ which is defined as
\bq\label{MBTr}
|\Phi^{(\varepsilon_1)}_{\theta} \rangle &\equiv& \frac{d}{d\theta}|\Psi^{(\varepsilon_1)}_{\theta}\rangle\nonumber\\
&=&-ie^{-i{\hat L}\theta}\sum_{{\cal L}>0}{\cal L}
\beta_{\cal L}\left(|{\cal L}\rangle-|-{\cal L}\rangle\right). 
\eq
The broken-symmetry state with the excitation to the translation mode is thus given by 
\bq
|\tilde{\Psi}_{\theta}^{(\varepsilon_1)}\rangle = \frac{1}{\sqrt{1+|\zeta|^2}}\left(|\Psi^{(\varepsilon_1)}_{\theta}\rangle+\frac{\zeta}{{\cal N}_{\rm tr}}|\Phi^{(\varepsilon_1)}_{\theta}\rangle\right), 
\eq
where $\zeta$ denotes the relative amplitude of the excitation, 
and the energy difference associated with the excitation of the translation mode is given by 
\bq\label{Etr}
E_{\rm tr}&\equiv& \langle \tilde{\Psi}_{\theta}^{(\varepsilon_1)}|\hat{K}_1|\tilde{\Psi}_{\theta}^{(\varepsilon_1)}\rangle-
\langle \Psi_{\theta}^{(\varepsilon_1)}|\hat{K}_1|\Psi_{\theta}^{(\varepsilon_1)}\rangle\nonumber\\
&=&\frac{|\zeta|^2}{1+|\zeta|^2}
\left( -\langle \Psi_{\theta}^{(\varepsilon_1)}|\hat{K}_1|\Psi_{\theta}^{(\varepsilon_1)}\rangle\right.\nonumber\\
&\ &\qquad\qquad\quad \left. +\frac{1}{|{\cal N}_{\rm tr}|^2}\langle \Phi_{\theta}^{(\varepsilon_1)}|\hat{K}_1|\Phi_{\theta}^{(\varepsilon_1)} \rangle\right), 
\eq
where ${\cal N}_{\rm tr}\equiv \sqrt{\langle \Phi_{\theta}^{(\varepsilon_1)}| \Phi_{\theta}^{(\varepsilon_1)}\rangle}$.
We note that $E_{\rm tr}$ becomes zero in the limit of $\zeta \to 0$. 

\subsubsection{Quantum fluctuations in position and momentum}\label{Pot_Fluct}

\begin{figure}[t]
\includegraphics[scale=0.55]{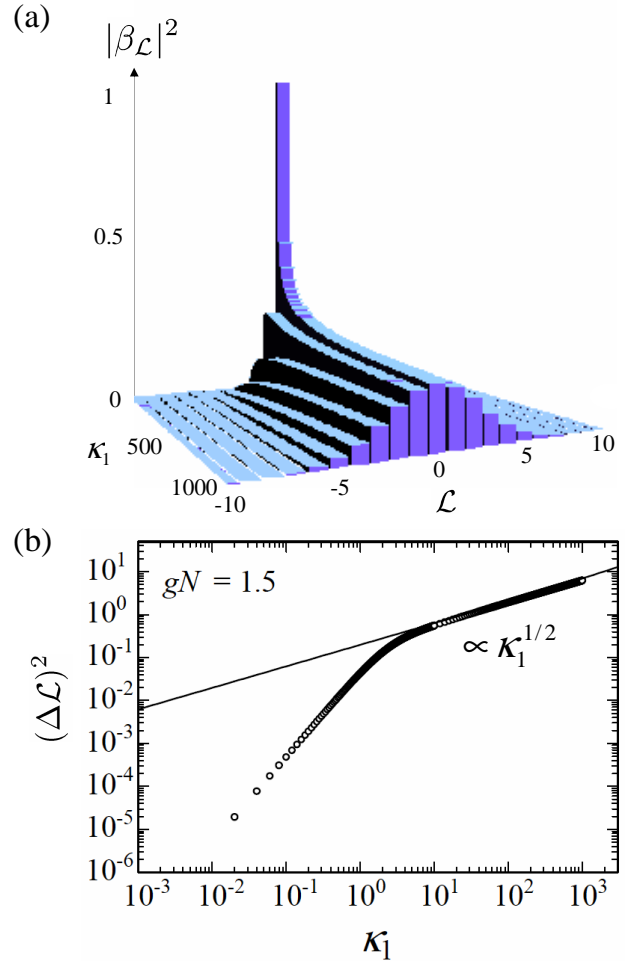}
\caption{
(Color online) (a) Distribution $|\beta_{\cal L}|^2$ of the angular momentum ${\cal L}$ 
in the many-body ground state $|\Psi^{(\varepsilon_1)}_{\theta}\rangle$ for $gN=1.5$. 
(b) Width of angular-momentum distribution 
in the many-body ground state $|\Psi^{(\varepsilon_1)}\rangle$ (open circles) and 
that in the Bogoliubov ground state $|\Psi^{\rm (B)}\rangle$ (solid line) given by 
Eq.~(\ref{P_width_Beta}).
}
\label{fig_pot1_fluct}
\end{figure}

Let us investigate the quantum fluctuations in the ground state 
$|\Psi_{\theta}^{(\varepsilon_1)}\rangle$ as a function of the 
magnitude of the symmetry-breaking potential. 
The angular-momentum distribution is given by 
$|\beta_{\cal L}|^2=|\langle {\cal L}|\Psi_{\theta}^{(\varepsilon_1)} \rangle|^2$, 
which becomes Gaussian-like for $\kappa_1 \gg 1$ as shown in Fig.~\ref{fig_pot1_fluct}(a). 
In Fig.~\ref{fig_pot1_fluct}(b) we plot the variance $(\Delta {\cal L})^2$ of 
the angular-momentum distribution with open circles. 
The deviation in $\kappa_1$ dependence of $(\Delta {\cal L})^2$ in 
$0\le \kappa_1 \lesssim 1$ arises because in this region 
the energy scale of the symmetry-breaking 
potential is smaller than the energy gap between 
the ground and the first excited states, 
and the angular-momentum fluctuation is suppressed. 
Once the matter wave is localized by the superposition of the form (\ref{MBLoc}), 
the fluctuation $\Delta{\cal L}$ of the 
angular-momentum distribution is proportional to $\kappa_1^{1/4}$. 

It follows from the uncertainty relation $\Delta{\cal L}\Delta{\theta}_{\rm c.m.}\simeq 1$ that 
the center-of-mass fluctuation of the localized state $|\Psi_{\theta}^{(\varepsilon_1)}\rangle$ 
for $\kappa_1 \gtrsim 1$ is given by 
\bq\label{p-cm-fluct}
\Delta \theta_{\rm c.m.} \simeq (\Delta {\cal L})^{-1}\propto \kappa_1^{-1/4}, 
\eq
which is significant for small perturbations~\cite{rotation}. 
In Sec.~\ref{Bog_pot1}, we will show that this $\kappa_1$ dependence is well 
described by the Bogoliubov ground state that 
takes into account this position fluctuation by the depletion of the condensate. 


\subsubsection{Condensate fraction}

\begin{figure}
\includegraphics[scale=0.48]{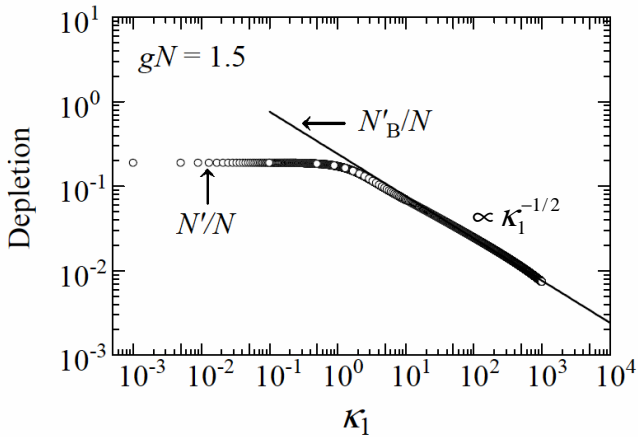}
\caption{Depletions of the condensate obtained by 
the diagonalization of the Hamiltonian (open circles), and 
by the Bogoliubov theory given in Eq.~(\ref{P_Bog_Dep}) (solid line). 
}
\label{fig_pot1_depl}
\end{figure}

By using the many-body ground state $|\Psi_{\theta}^{(\varepsilon_1)}\rangle$ in the presence of 
the symmetry-breaking potential, and diagonalizing the single-particle 
density matrix 
$\rho(l,l')=\langle \Psi_{\theta}^{(\varepsilon_1)}|
\hat{c}_{l'}^{\dagger}\hat{c}_l 
|\Psi_{\theta}^{(\varepsilon_1)}\rangle$, 
we obtain the three eigenvalues in the truncated bases. 
We define the number of depleted atoms $N'$ of 
the many-body ground state as $N'/N \equiv 1-\lambda^{(\varepsilon_1)}$, where 
$\lambda^{(\varepsilon_1)}$ is the maximum eigenvalue of $\rho$. 
We plot $N'/N$ in Fig.~\ref{fig_pot1_depl} as a function of $\kappa_1$. 
For $\kappa_1=0$, the maximum depletion is obtained due to 
the fragmentation of the many-body ground state, and 
it remains constant for $\kappa_1 \lesssim 1$. 
The depletion $N'$ suddenly begins to decrease at $\kappa_1 \simeq 1$, and 
the $\kappa_1$ dependence for $\kappa_1 \gtrsim 1$ is found to be $\kappa_1^{-1/2}$. 
The condensate fraction thus increases when the ground state begins to 
localize. 

\subsection{Case of a double-well potential}\label{MB_pot2}

We next discuss what happens when there exists a symmetric double-well potential of the form 
$V_2(\theta)=\cos 2\theta$, 
and discuss the ground state of the Hamiltonian 
\bq
\hat{K}_2=\hat{H}+\varepsilon_2\hat{V}_2,
\eq
where 
\bq
\hat{V}_2=\int_0^{2\pi} d\theta\hat{\psi}^{\dagger}(\theta)V_2(\theta)\hat{\psi}(\theta). 
\eq
In the double-well potential, a rich variety of experiments, e.g., 
diagnosing the correlation functions from interference and 
uncertainty relations, become possible~\cite{Orzel,DW}. 

\begin{figure}[t]
\includegraphics[scale=0.48]{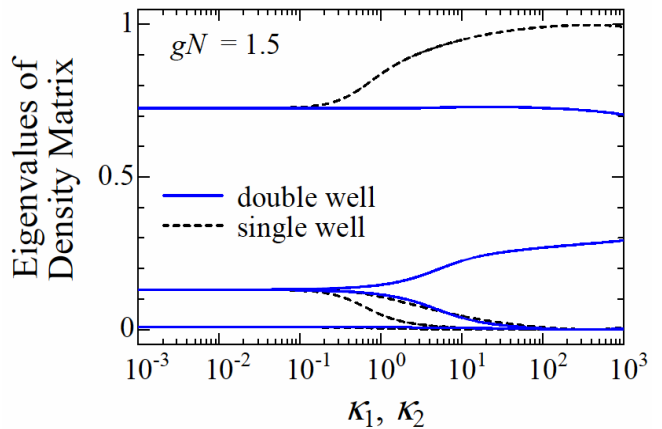}
\caption{
(Color online) Eigenvalues of the reduced single-particle density matrix 
for the double-well potential (solid curves) 
as a function of $\kappa_2 \equiv \varepsilon_2 N^2$. 
Dashed curves show the 
results obtained for a single-well potential as a function of $\kappa_1\equiv \varepsilon_1 N^2$. 
In the latter case only one large eigenvalue survives for $\kappa_1 \gtrsim 1$. 
}
\label{fig_pot2_cf}
\end{figure}

For an attractive condensate in a sufficiently deep symmetric double-well potential, 
the ground state may form a Schr\"{o}dinger's cat state, 
where the localized macroscopic states on the left and the right wells are 
in a superposition state~\cite{HC,CLMZ,SC,MY}. 
We diagonalize the Hamiltonian $\hat{K}_2$ including 
angular-momentum states $l=0,\pm 1, \pm 2$ because the symmetry of 
the potential requires the inclusion of the $l=\pm 2$ states. 
The solid curves in Fig.~\ref{fig_pot2_cf} show eigenvalues of the reduced 
single-particle density matrix. 
The presence of more than one large eigenvalue is a signature 
of the Schr\"{o}dinger's cat state. 
While the maximum eigenvalue $\lambda_1$ increases with increasing $\varepsilon_1$, 
it decreases slowly with increasing $\varepsilon_2$, and 
simultaneously the second maximum eigenvalue $\lambda_2$ 
begins to grow. These eigenvalues $\lambda_{1,2}$ eventually approach $1/2$, corresponding 
to the states localized on the one or the other well. 

\begin{figure}
\includegraphics[scale=0.45]{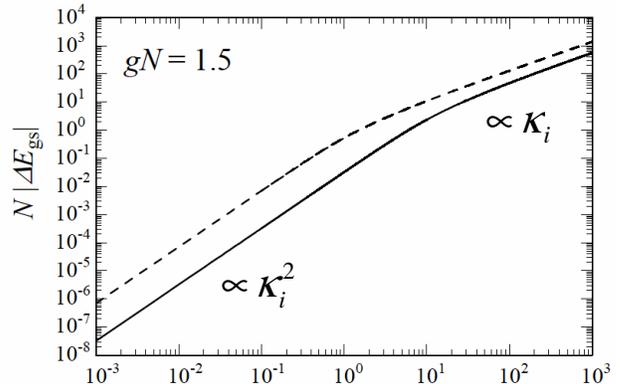}
\caption{Ground-state energy as a function of $\kappa_2$ (solid curve) in the presence of 
a double-well potential measured from $E_0=\langle 0 | \hat{H} | 0\rangle$. 
Dashed curve shows the results of a single-well potential. 
}
\label{fig_pot2_ene}
\end{figure}
 
In the case of the double-well potential, 
the many-body ground state is also described by the superposition in 
the form of Eq.~(\ref{MBLoc}). 
Figure~\ref{fig_pot2_ene} shows that the $\kappa_2$ dependence of the energy difference 
$\langle \Psi^{(\varepsilon_2)}_{\theta}|\hat{K}_2|\Psi^{(\varepsilon_2)}_{\theta} \rangle-\langle 0|\hat{H}|0\rangle$ 
is similar to that of the single-potential case shown as a dashed curve. 
This indicates that the superposition begins near $\kappa_2\simeq 1$, since 
the energy at that point is of the order $1/N$. 


\subsection{Quantum measurement}\label{MB_QM}

We next consider quantum fluctuations caused by 
repeated measurements. 
The issue of interference of the two independent BECs 
was discussed from the viewpoint of quantum measurements in Refs.~\cite{JY,CD97}. 
Below we discuss an analogous process in which 
a fragmented condensate makes a transition to 
a single condensate via a quantum measurement~\cite{KSU,MA}. 

\subsubsection{Quantum fluctuations in position and momentum}

The action of the quantum measurement at the position $\theta_j$ relates 
the postmeasurement state $|\Psi^{(j)}\rangle$ to the 
premeasurement state $|\Psi^{(j-1)}\rangle$ as 
\begin{eqnarray}\label{action}
|\Psi^{(j)}\rangle = 
\frac{\hat{\psi}(\theta _j)|\Psi^{(j-1)}\rangle}{\sqrt{\langle \Psi^{(j-1)}|\hat{\psi}^{\dagger}(\theta_j)\hat{\psi}(\theta_j)|\Psi^{(j-1)}\rangle}}, 
\end{eqnarray}
where $j$ is the number of measurements, and the initial state $|\Psi^{(j=0)}\rangle$ is taken to be the 
many-body ground state. 
The subsequent measurement position $\theta_{j+1}$ after the $j$th measurement 
is probabilistically determined according to the density 
distribution $n_j(\theta)=N_j\rho^{(j)}(\theta,\theta)$ of the premeasurement state, 
where $N_j=N_{\rm init}-j$ is the number of atoms. 

\begin{figure}[t]
\includegraphics[scale=0.55]{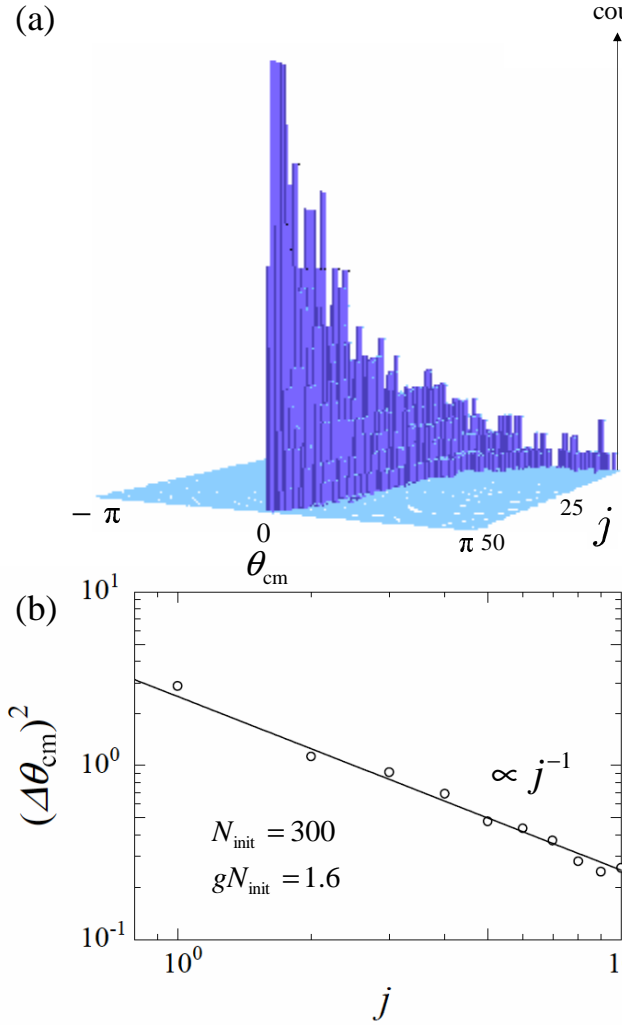}
\caption{
(Color online) (a) Cumulative number of counts for the center-of-mass position obtained from 
2000 runs of independent simulation of quantum measurements for $gN_{\rm init}=1.6$ 
and $N_{\rm init}=300$. 
(b) Center-of-mass fluctuations obtained 
by the numerical simulations (open circles). 
The solid straight line obeys $(\Delta \theta_{\rm c.m.})^2\propto j^{-1}$ and 
is drawn as a guide to the eye. 
}
\label{fig_mea_cm}
\end{figure}

\begin{figure}[t]
\includegraphics[scale=0.55]{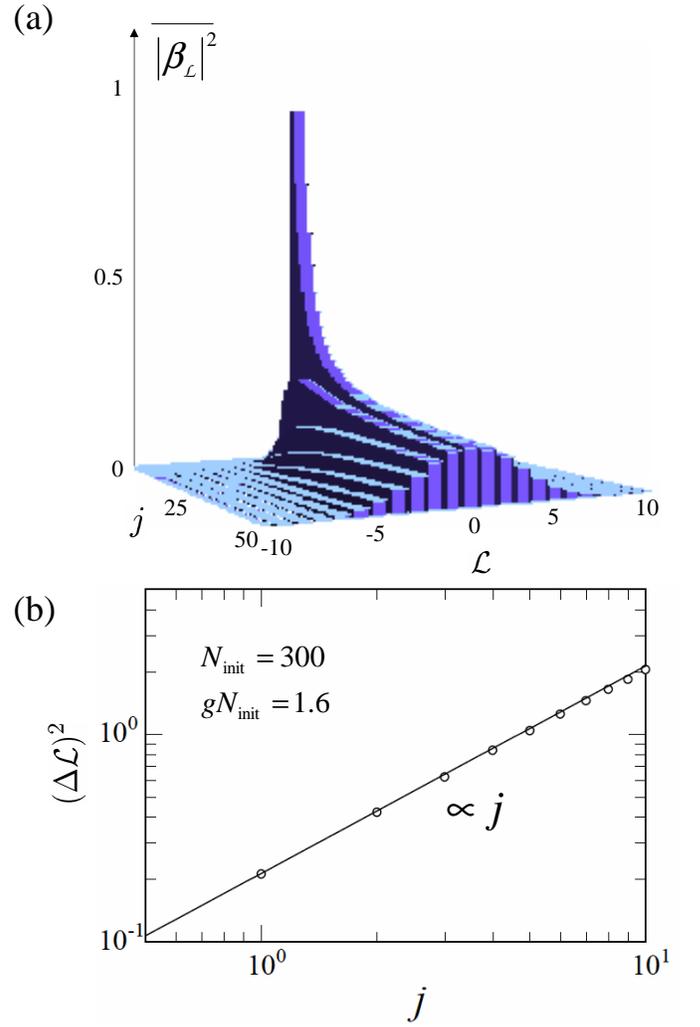}
\caption{
(Color online) (a) Ensemble average of the angular-momentum distribution 
$|{\beta}_{\cal L}|^2$ 
obtained from 2000 runs of the numerical simulation, 
and (b) its width (open circles) for $gN_{\rm init}=1.6$ and $N_{\rm init}=300$. 
The solid line is given by Eq.~(\ref{M_Beta}). 
}
\label{fig_mea_am}
\end{figure}

We perform 2000 runs of the numerical simulation of 50 quantum measurements (\ref{action}) 
($0\le j \le 50$) starting from the ground state with $N_{\rm init}$ atoms. 
By the measurements, the density distribution $n_j(\theta)$ is found to localize, 
and we denote the center-of-mass position of $n_j(\theta)$ by $\theta_{\rm c.m.}^{(j)}$. 
We find that as $j$ increases, $\theta_{\rm c.m.}^{(j)}$ 
converges to a certain position, which is random from 
run to run. 
We therefore rotate the system so that $\theta_{\rm c.m.}^{(j=50)}$ is located at the origin, and plot the distribution 
of $\theta_{\rm c.m.}^{(j)}$ for 2000 runs of measurements in Fig.~\ref{fig_mea_cm}(a). 
Hence, this distribution represents the center-of-mass fluctuation after 
the $j$th measurement. 
The variance of the distribution is plotted in Fig.~\ref{fig_mea_cm}(b), 
which shows that 
\bq
(\Delta\theta)^2 \propto 1/j. 
\eq
Because of the translation symmetry of the initial state $|\Psi^{(j=0)}\rangle$, 
the center-of-mass fluctuation is maximal before the measurement. 
The repeated measurement process reduces the position fluctuation, and 
eventually projects the translation-invariant state $|\Psi^{(j=0)}\rangle$ 
onto a broken-symmetry state whose center-of-mass localizes at a certain position. 

Because the quantum measurement is a stochastic process, we consider 
the ensemble average of $|\beta_{\cal L}|^2=|\langle {\cal L}|\Psi^{(j)}\rangle|^2$ 
over the 2000 runs of independent simulations. The result is shown 
in Fig.~\ref{fig_mea_am}(a). We find from Fig.~\ref{fig_mea_am}(b) that the distribution of 
the ensemble-averaged angular momentum obeys the Gaussian with its width given by 
\bq
(\Delta {\cal L})^2 \propto j. 
\eq 
In the present 1D system, the center-of-mass position 
$\theta_{\rm c.m.}=\sum_{k=1}^{N}\theta_k/N$ and the angular momentum 
${\cal L}=\sum_{k=1}^N (-i \partial)/(\partial\theta_k)$ 
obey the commutation relation
\bq
\left[\theta_{\rm c.m.},{\cal L}\right]=i. 
\eq
Their fluctuations 
$\Delta\theta_{\rm c.m.} \equiv \theta_{\rm c.m.}-\langle \theta_{\rm c.m.} \rangle$ 
and $\Delta {\cal L} \equiv {\cal L}-\langle {\cal L} \rangle$ therefore obey
\bq
\Delta\theta_{\rm c.m.}\Delta{\cal L} \ge \frac{1}{2}. 
\eq
From Figs.~\ref{fig_mea_cm} and \ref{fig_mea_am}, we found the uncertainty 
\bq\label{mea_uncertainty}
\Delta\theta_{\rm c.m.}\Delta{\cal L} \simeq 0.73, 
\eq
which is larger than that of the minimum uncertainty state. 


\subsubsection{Condensate fraction}

\begin{figure}[t]
\includegraphics[scale=0.45]{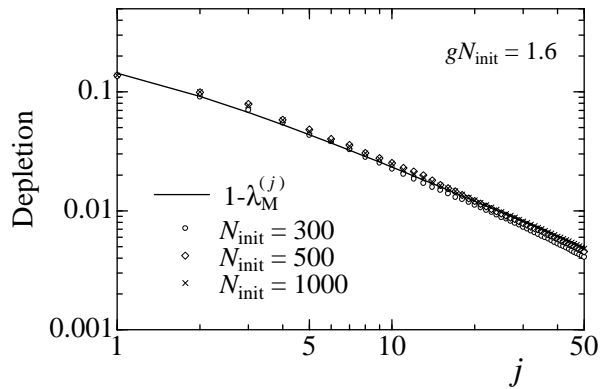}
\caption{
Depletion of the condensate for $gN_{\rm init}=1.6$.
Symbols ($\circ, \diamond$,{\footnotesize $\times$}) show the results of numerical simulations, 
and the solid curve shows a theoretical one in Eq.~(\ref{M_CF}), 
where $j$ is the number of measurements. 
}
\label{fig_mea_depl}
\end{figure}

We study the change in the condensate fraction 
as a function of the number of measurements $j$. 
In Fig.~\ref{fig_mea_depl}, we show the ensemble-averaged depleted fraction 
of the condensate in the state $|\Psi^{(j)}\rangle$ for three different values 
of the initial number of atoms $N_{\rm init}$. 
The depletion decreases monotonically with increasing $j$. 
The change in the condensate fraction with respect to the number of measurements $j$ 
is not sensitive to the initial number of atoms $N_{\rm init}$, and 
is determined by $gN_{\rm init}$. 
Like a symmetry-breaking potential, the repeated quantum measurements 
change the fragmented condensate having a translation symmetry into 
a translation-symmetry broken single condensate with 
reduced center-of-mass fluctuations. 

\section{Bogoliubov and Extended Mean-Field Theories}\label{MFBog}

In the previous sections we showed that center-of-mass fluctuations are significant when 
the symmetry-breaking perturbations are sufficiently small. 
We here analytically treat the many-body 
state with large center-of-mass fluctuations 
for both cases of the symmetry-breaking potential and the quantum measurement. 

\subsection{Mean-field ground state in a homogeneous ring}\label{GPuniform}

By replacing the field operator with a $c$-number wave function as 
$\hat{\psi}(\theta)\to \phi(\theta)$, 
we obtain GP energy functional
\bq\label{GPene}
E^{\rm (GP)}_0[\phi]=N\int_0^{2\pi}d\theta(|\nabla\phi|^2-\pi gN|\phi|^4). 
\eq
We employ a variational wave function 
\bq\label{GPvar}
\phi(\theta)=\frac{1}{\sqrt{2\pi}}(\alpha_0+\alpha_1e^{i\theta}+\alpha_{-1}e^{-i\theta}), 
\eq
where $\alpha_{0,\pm 1}$ are complex variational parameters which obey 
the normalization condition $|\alpha_{-1}|^2+|\alpha_0|^2+|\alpha_1|^2=1$ 
and are to be determined so as to minimize $E_0^{\rm (GP)}$. 
For $gN < 1$, $E_0^{\rm (GP)}$ is minimized when $\alpha_0=1, \alpha_{\pm 1}=0$, 
i.e., the condensate wave function is uniform, $\phi(\theta)=1/\sqrt{2\pi}$. 

For $gN \ge 1$, $E_0^{\rm (GP)}$ is minimized when 
\bq\label{GPa}
\alpha_0=\sqrt{\frac{3gN+2}{7gN}},\ 
\alpha_1=\alpha_{-1}^*=e^{-i\theta_{\rm c.m.}}\sqrt{\frac{2(gN-1)}{7gN}}, 
\eq
where the global phase is chosen so that $\alpha_0$ is 
real without loss of generality. 
The corresponding wave function 
\bq\label{GPsol}
\phi(\theta-\theta_{\rm c.m.})=\frac{1}{\sqrt{2\pi}}\left[\alpha_0+2\alpha_1\cos(\theta-\theta_{\rm c.m.})\right], 
\eq
is a spontaneously broken-symmetry state. 
Note that all states described by Eq.~(\ref{GPsol}) are 
degenerate with respect to an arbitrary center-of-mass coordinate $\theta_{\rm c.m.}$. 
The critical point $gN=1$ of the quantum phase transition 
obtained by the present mean-field theory agrees with 
the exact analysis of the 1D GP equation, and Eq.~(\ref{GPa}) indeed corresponds to 
the dominant coefficients of the plane-wave expansion of 
the exact solution of the 1D GP equation for $gN \ge 1$~\cite{Carr,2003}, 
\bq
\phi(\theta-\theta_{\rm c.m.})
&\propto&
{\rm dn}\left(\left.\frac{K(m)}{\pi}(\theta-\theta_{\rm c.m.})\right|m\right)\nonumber\\
&=&
1+\sum_{l=1}^{\infty}\frac{e^{il(\theta-\theta_{\rm c.m.})}+e^{-il(\theta-\theta_{\rm c.m.})}}{{\rm cosh}(\pi lK'/K)}, 
\eq
where dn is one of the Jacobian elliptic functions, 
$K(m)\equiv \int_0^{\pi/2}dx/\sqrt{1-m \sin^2 x}$ is the complete elliptic integral of the first kind, 
and $0 \le m \le 1$. 
From the definition of the GP mean-field theory, 
all atoms occupy the same single-particle state $\phi$, and 
the condensate fraction is unity for all $gN$. 
For $gN \ge 1$, the ground state of the GP equation given in Eq.~(\ref{GPsol}) 
is thus qualitatively different 
from the many-body ground state $|0\rangle_{\sigma=0}$ in both 
the condensate fraction and the translation symmetry. 


\subsection{Infinitesimal potential}\label{Bog_pot1}


\subsubsection{Gross-Pitaevskii and Bogoliubov analyses}\label{GPBana}

We introduce the symmetry-breaking potential $V_1(\theta)$ which is assumed to 
be infinitesimal, i.e., of the order of $1/N$. 
The GP energy functional in the presence of a symmetry-breaking potential 
$V_1(\theta)=\cos\theta$ is given by 
\bq\label{P_GPene}
E_1^{\rm (GP)}[\phi]=N\!\int_0^{2\pi}\!\!\!d\theta 
\left[|\nabla\phi(\theta)|^2+\varepsilon_1 V_1(\theta-\pi)|\phi(\theta)|^2\right.\nonumber\\
\left.-\pi gN|\phi(\theta)|^4\right]. 
\eq
When we employ the same variational function of the form in Eq.~(\ref{GPvar}), 
$E^{\rm (GP)}_1$ is minimized when the variational coefficients $\alpha_0$ and $\alpha_1=\alpha_{-1}$ 
are given by 
\bq
\alpha_1^2\!\!&=&\!\!\frac{2(gN-1)}{7gN}+\frac{8-gN}{7gN\sqrt{2(gN-1)(3gN+4)}}\varepsilon_1+O(\varepsilon_1^2),\label{P_GPa1}\nonumber\\
\ \ \\
\alpha_0^2\!\!&=&\!\!1-2 \alpha_1^2, \label{P_GPa0}
\eq
where we assume that the center-of-mass position is located at $\theta_{\rm c.m.}=0$. 
The corresponding chemical potential and energy read 
\bq
\mu\!\!&=&\!\!\frac{8-15 gN}{14}-\frac{1}{7gN}
\left[\frac{\sqrt{2}(8-gN)}{\sqrt{(gN-1)(3gN+4)}}\right.\nonumber\\
&\ &\left.+\sqrt{2(gN-1)(3gN+4)}\right]\varepsilon_1+O(\varepsilon_1^2),\label{P_GPmu}\\
E_1^{\rm (GP)}/N\!\!\!&=\!\!\!&\frac{1}{28gN}\left(-15g^2N^2+16gN-8\right)\nonumber\\
&\ +&\!\!\frac{2\sqrt{2}}{7gN}\sqrt{(gN-1)(3gN+4)}\ \varepsilon_1+O(\varepsilon_1^2).\label{P_GPene}
\eq
From Eq.~(\ref{P_GPene}) we obtain 
\bq\label{GPene_dif}
N(E_1^{\rm (GP)}-E_0^{\rm (GP)})\hspace{5cm}\nonumber\\
=\frac{2}{7gN}\sqrt{2(gN-1)(3gN+4)}\ \kappa_1+O(\varepsilon_1^2), 
\eq
which agrees with the $\kappa_1$ dependence of $N\Delta E_{\rm gs}$ 
[see the dashed curve in Fig.~\ref{fig_pot1_ene}]. 

In the case of uniform space ($\varepsilon_1= 0$), 
the translation mode can be excited without energy cost for $gN\gtrsim 1$~\cite{2003}, 
which corresponds to the Nambu-Goldstone mode. 
In the presence of a nonuniform potential, the translation mode slightly costs energy. 
By solving the Bogoliubov-de Gennes equations
\bq
\left[\!
\begin{array}{cc}
-\partial_{\theta}^2-\varepsilon_1V_1(\theta)-\mu-2\pi gN|\phi|^2 \  &\  -\pi gN \phi^2\\
\partial_{\theta}^2+\varepsilon_1V_1(\theta)+\mu+2\pi gN|\phi|^2  \  &\  \pi gN {\phi^{*}}^2
\end{array}
\!\right]\!\!
\left[\!
\begin{array}{c}
u\\
v
\end{array}
\!\right]\!\!
=\!\!\Lambda
\!\left[\!
\begin{array}{c}
u\\
v
\end{array}
\!\right],\nonumber
\ \!\!\!\!\!\!\!\!\!\!\!\!\\
\eq
we obtain the eigenvector corresponding to the lowest-energy translation mode as 
\bq
\left[
\begin{array}{c}
u\\
v
\end{array}
\right]
&\propto&
\sin\theta\left[
\begin{array}{c}
1+\sqrt{\delta /2}-\delta/4+O(\varepsilon^{3/2})\\
1-\sqrt{\delta /2}-\delta/4+O(\varepsilon^{3/2})\\
\end{array}
\right],\label{BogUV}\\
\Lambda&=&\pm \sqrt{\delta^2+2\delta},\label{BogLambda} 
\eq
where $\delta$ is given by 
\bq\label{delta}
\delta\equiv \frac{7\varepsilon_1}{2+5gN}\sqrt{\frac{3gN+4}{2gN-2}}. 
\eq
The corresponding quasiparticle operator is given by 
\bq\label{opb}
\hat{b}\propto u\hat{T}+v\hat{T}^{\dagger},
\eq
where 
\bq
\hat{T}=\frac{1}{\sqrt{2}}(\hat{c}_1-\hat{c}_1^{\dagger}). 
\eq
The GP ground state whose center-of-mass position is located at 
$\theta_{\rm c.m.}$ is written as 
\bq\label{cmGP}
|\Psi^{\rm (GP)}_{\theta_{\rm c.m.}}\rangle=\frac{1}{\sqrt{N!}}
\left[\alpha_0\hat{c}_0^{\dagger}+\alpha_1(\hat{c}_1^{\dagger}e^{i\theta_{\rm c.m.}}+\hat{c}_{-1}^{\dagger}e^{-i\theta_{\rm c.m.}})\right]^N|{\rm vac}\rangle, \nonumber\\
\!\!\!
\eq
and the derivative with respect to $\theta_{\rm c.m.}$ gives 
\bq
\left.\frac{\partial}{\partial \theta_{\rm c.m.}}|\Psi^{\rm (GP)}_{\theta_{\rm c.m.}}\rangle\right|_{\theta_{\rm c.m.}=0}
\propto \hat{T}|\Psi^{\rm (GP)}_{\theta_{\rm c.m.}=0}\rangle. 
\eq
Thus, the operator $\hat{T}$ serves as the generator of the translation of 
the GP ground state $|\Psi^{\rm (GP)}\rangle$. 


\subsubsection{Quantum fluctuations in position and momentum}

We consider the Bogoliubov ground state. 
Since the translation costs little energy, the center-of-mass fluctuation is 
expected to be dominant in the Bogoliubov ground state. 
We therefore take into account the translation mode only. 
The Bogoliubov quasiparticle operator~(\ref{opb}) is written as 
\bq
\hat{b}=
e^{-\frac{r}{2}(\hat{T}^{\dagger 2}-\hat{T}^2)}
\hat{T}
e^{\frac{r}{2}(\hat{T}^{\dagger 2}-\hat{T}^2)}
=\hat{T}\cosh r+\hat{T}^{\dagger}\sinh r, 
\eq
where the parameter $r$ is related to the Bogoliubov 
quasiparticle and hole amplitudes $u$ and $v$ by 
$(\cosh r,\sinh r)\propto (u,v)$. 
We evaluate the angular-momentum distribution in the Bogoliubov ground state
\bq\label{Boggs}
|\Psi^{\rm (B)}\rangle\!\!&=&\!\!e^{\frac{r}{2}(\hat{T}^{\dagger 2}-\hat{T}^2)}|\Psi^{\rm (GP)}_{\theta_{\rm c.m.}=0}\rangle\nonumber\\
&=&\!\!e^{\frac{r}{2}(\hat{T}^{\dagger 2}-\hat{T}^2)}\frac{1}{\sqrt{N!}}\!(\alpha_0\hat{c}_0^{\dagger}+\sqrt{2}\alpha_1\hat{S}^{\dagger})^N\!|{\rm vac}\rangle, 
\eq
where
\bq 
\hat{S}\equiv \frac{1}{\sqrt{2}}(\hat{c}_1+\hat{c}_{-1}). 
\eq
Let us approximate the angular-momentum operator 
\bq
\hat{L}=\hat{n}_1-\hat{n}_{-1}=\hat{S}^{\dagger}\hat{T}+\hat{S}\hat{T}^{\dagger},
\eq
as 
\bq
\hat{L}\to \hat{\cal L}\equiv \sqrt{2N}\alpha_1(\hat{T}+\hat{T}^{\dagger}),
\eq
by the replacement of $\hat{S}$ with its mean-field value $\sqrt{2N}\alpha_1$ 
under the assumption that the ratio of the condensate atoms to the excited atoms is sufficiently large. 
By using the relation (see Appendix~\ref{adx} for proof)
\bq\label{Op}
e^{-\frac{r}{2}(\hat{T}^{\dagger 2}-\hat{T}^2)}|{\rm vac}\rangle
=\sqrt{\frac{1+\eta}{\cosh r}}e^{-\frac{\eta}{2}(\hat{T}+\hat{T}^{\dagger})^2}|{\rm vac}\rangle, 
\eq
we obtain the Bogoliubov  ground state as
\bq\label{P_Bgs}
|\Psi^{\rm (B)}\rangle\simeq {\cal N}_{\rm B}\exp\left(-\frac{\eta\hat{\cal L}^2}{4N\alpha_1^2}\right)
|\Psi^{\rm (GP)}_{\theta_{\rm c.m.}=0}\rangle,
\eq
where ${\cal N}_{\rm B}$ is the normalization constant, and 
\bq
\eta\!\!&=&\!\!\!\frac{\tanh r}{1-\tanh r}=\frac{v/u}{1-v/u}\nonumber\\
&=&\!\!\!\sqrt{\frac{5 gN+2}{14}}\!\left(\frac{2gN-2}{3gN+4}\right)^{1/4}\!\!\!\varepsilon_1^{-1/2}\!-\!\frac{1}{2}\!+\!O(\varepsilon_1^{1/2}). 
\eq
We thus find that the angular-momentum distribution in the Bogoliubov ground state is given by 
\bq\label{P_Beta}
|\langle {\cal L}|\Psi^{\rm (B)}\rangle|^2&=&
\frac{1}{\sqrt{2 \pi (\Delta{\cal L}^{\rm (B)})^2}}e^{-\frac{{\cal L}^2}{2 (\Delta {\cal L}^{\rm (B)})^2}},
\eq
where the width of this distribution is calculated to be (see Appendix~\ref{AM_fluct})
\bq\label{P_width_Beta}
(\Delta {\cal L}^{\rm (B)})^2
&=&N\alpha_1^2/\eta\nonumber\\
&=&\frac{2gN-2}{7gN}\left(\sqrt{2F}+F\sqrt{\varepsilon_1}\right)\sqrt{\kappa_1}
\propto \kappa_1^{1/2},\nonumber\\
\eq
with 
\bq
F\equiv \frac{7}{5 gN+2}\sqrt{\frac{3gN+4}{2gN-2}}.
\eq
The distribution of the angular momentum in $|\Psi^{\rm (B)}\rangle$ 
is in excellent agreement with the distribution coefficient 
$|\beta_{\cal L}|^2= |\langle {\cal L}|\Psi^{(\varepsilon_1)}_{\theta}\rangle|^2$ in 
the many-body ground state~(\ref{MBLoc}) 
which is obtained numerically by the diagonalization of 
the Hamiltonian $\hat{K}$ in Eq.~(\ref{hamK}). 


\subsubsection{Condensate fraction}

The condensate fraction of the Bogoliubov ground state is less than unity 
because of virtual particle-pair excitations mainly to the translation mode. 
The virtual excitations are the physical origin of 
quantum fluctuations in the position and momentum described above. 
The number of depleted atoms is calculated from Eq.~(\ref{BogUV}) as 
\bq\label{P_Bog_Dep}
N'_{\rm B}= \int_0^{2\pi} v^2(\theta)d\theta=\frac{1}{\sqrt{8\delta}}-\frac{1}{2}+O(\sqrt{\delta}), 
\eq
where $\delta$ is given in Eq.~(\ref{delta}). 
The result is shown as the solid line in Fig.~\ref{fig_pot1_depl} 
which agrees well with the results of the exact diagonalization for $\kappa_1 \gtrsim 1$. 

We thus find that in the presence of the symmetry-breaking potential, 
the Bogoliubov ground state very well reproduces 
the energy, the angular-momentum distribution, and the depletion of the 
condensate obtained by the exact diagonalization for $\kappa_1 \gtrsim 1$. 
This is because the virtual excitations described by the Bogoliubov ground state 
lead to the center-of-mass fluctuations above the mean-field ground state. 


\subsection{Quantum measurement}\label{MF_QM}


\subsubsection{Quantum fluctuations in position and momentum}

In this section the suppression of the center-of-mass fluctuation 
via repeated quantum measurements shown in Fig.~\ref{fig_mea_cm} is 
investigated semiclassically by generalizing mean-field theory. 
We assume as an initial state superposition state of the GP solution 
$|\Psi_{\theta_{\rm c.m.}}^{\rm (GP)}\rangle$ 
with respect to the center-of-mass position $\theta_{\rm c.m.}$, 
\bq\label{psicl}
|\Psi_{\rm cl}^{(j)}\rangle &\equiv& \int d\theta_{\rm c.m.}A_j(\theta_{\rm c.m.})|\Psi_{\theta_{\rm c.m.}}^{\rm (GP)}\rangle. 
\eq
We assume that for $j=0$, $A_0(\theta_{\rm c.m.})$ is a constant. 
The repeated measurements would select a localized soliton 
at a certain $\theta_{\rm c.m.}$, 
$\phi(\theta-\theta_{\rm c.m.})=\left[\alpha_0+2\alpha_1\cos(\theta-\theta_{\rm c.m.})\right]/\sqrt{2\pi}$
with probability $A^2_j(\theta_{\rm c.m.})$. 
If the first atom is detected at a position $\theta_1$, 
the postmeasurement distribution may be approximated by 
\bq
A_1 (\theta_{\rm c.m.})=\frac{1}{\sqrt{2\pi}}\left[\alpha_0+2 \alpha_1 \cos(\theta_1-\theta_{\rm c.m.})\right], 
\eq
where $\alpha_0$ and $\alpha_1$ are given by Eqs.~(\ref{GPa}). 
Likewise, after the $j$th measurement, the distribution becomes 
\bq\label{calP}
A_j(\theta_{\rm c.m.})\propto \prod_{i=0}^{j}
\frac{1}{\sqrt{2\pi}}\left[\alpha_0+2\alpha_1\cos(\theta_i-\theta_{\rm c.m.})\right]. 
\eq
If we rotate the system so that $\theta_{\rm c.m.}\to 0$ for $j\to \infty$ (corresponding to the case of Fig.~\ref{fig_mea_cm}), 
the distribution of $\theta_i$ should become $[\alpha_0+2\alpha_1 \cos\theta]^2/(2\pi)$. 
Equation~(\ref{calP}) is then simplified for $j\gg 1$ as 
\bq
\ln A_j\!\!\!\!&(&\!\!\!\!\theta_{\rm c.m.})
=\sum_{i=0}^{j} \ln \left[\alpha_0+2\alpha_1\cos(\theta_i-\theta_{\rm c.m.})\right]+{\rm const}\nonumber\\
&\simeq&\! j\int d\theta n(\theta)\ln \left[\alpha_0+2\alpha_1\cos(\theta_i-\theta_{\rm c.m.})\right]+{\rm const}\nonumber\\
&\simeq&\!-2j\alpha_1^2\theta_{\rm c.m.}^2+{\rm const}+\cdots. 
\eq
The distribution function of the center-of-mass position $\theta_{\rm c.m.}$ 
can thus be approximated as 
\bq\label{Pj}
A_j(\theta_{\rm c.m.})\propto e^{-2 j \alpha_1^2 \theta_{\rm c.m.}^2}. 
\eq
It follows then that the center-of-mass position fluctuates according to
\bq\label{M_cm}
\Delta\theta_{\rm c.m.}\propto \frac{1}{\sqrt{j}}. 
\eq

The angular-momentum distribution of the state (\ref{psicl}), 
can also be analytically calculated. 
Substituting Eq.~(\ref{Pj}) in Eq.~(\ref{psicl}), 
we find that the angular-momentum distribution function is given by 
\bq
|\langle {\cal L}|\Psi_{\rm cl}^{(j)}\rangle|^2
\propto \frac{1}{\sqrt{2\pi (\Delta{\cal L})^2}}e^{-\frac{{\cal L}^2}{2(\Delta{\cal L})^2}},
\eq
with the width 
\bq\label{M_Beta}
(\Delta{\cal L})^2=2 \alpha_1^2 j. 
\eq
It follows from Eqs.~(\ref{M_cm}) and (\ref{M_Beta}) that during the repeated measurements 
the fluctuations in the center of mass and the angular momentum are found to 
obey the uncertainty relation 
\bq
\Delta\theta_{\rm c.m.}\Delta{\cal L}\simeq 1. 
\eq
Both Eqs.~(\ref{M_cm}) and (\ref{M_Beta}) are in excellent agreement with the 
ensemble averages of the corresponding results obtained by 
the numerical simulation as shown in Figs.~\ref{fig_mea_cm} and \ref{fig_mea_am}. 


\subsubsection{Condensate fraction}

We showed in the previous section that the repeated measurements suppress the center-of-mass fluctuations. 
We show here that this leads to an enhancement of the condensate fraction. 
The reduced single-particle density matrix of $|\Psi^{(j)}_{\rm cl}\rangle$ is obtained from 
Eqs.~(\ref{cmGP}), (\ref{psicl}), and (\ref{Pj}) as 
\bq\label{rho}
\rho(l,l')=\langle \hat{c}_{l'}^{\dagger}\hat{c}_l \rangle \qquad\qquad\qquad\qquad\qquad\qquad\qquad \nonumber\\
\to \left[
\begin{array}{ccc}
\alpha_1^2 
& \alpha_0\alpha_1 e^{-\nu/2}
& \alpha_1^2 e^{-2\nu}\\
\alpha_0\alpha_1 e^{-\nu/2}
& \alpha_0^2
& \alpha_0\alpha_1 e^{-\nu/2}\\
\alpha_1^2 e^{-2\nu}
& \alpha_0\alpha_1 e^{-\nu/2}
& \alpha_1^2
\end{array}
\right],
\eq\label{m-rho}
\!\!where $\nu\equiv 1/(8\alpha_1^2 j)$, and we assumed
\bq
\langle \Psi_{\theta'_{\rm c.m.}}|\Psi_{\theta_{\rm c.m.}} \rangle \simeq \delta(\theta_{\rm c.m.}'-\theta_{\rm c.m.}), 
\eq
for simplicity. 
The largest eigenvalue $\lambda_{\rm M}^{(j)}$ of $\rho(l,l')$ is given by 
\bq\label{M_CF}
&\lambda&\!\!\!\!_{\rm M}^{(j)}=\frac{1}{2}
\left[1-\alpha_1^2+\alpha_1^2 e^{-2\nu}\qquad\qquad\qquad\qquad\qquad\qquad\right.\nonumber\\
&+&\!\!\!\!\left.\sqrt{\!(1-\alpha_1^2+\alpha_1^2 e^{-2\nu})^2\!-\!4\alpha_1^2(1-2\alpha_1^2)(1-e^{-\nu})^2}\right]. 
\eq
Figure~\ref{fig_mea_depl} compares Eq.~(\ref{M_CF}) with 
the results obtained by numerical diagonalization, 
where the quantitative agreement is found. 
Like a symmetry-breaking potential, the repeated quantum measurements also change 
the fragmented condensate with a translation symmetry to a single condensate 
with a broken translation symmetry. 
In terms of the generalized mean-field theory developed in this section, 
the depletion of atoms is described by 
the amplitude $A_j$ in the state~(\ref{psicl}), the absolute square of 
which represents the semiclassical distribution function of the center of mass of the bright soliton. 


\section{Summary and Conclusions}\label{Sec5}

We have investigated the critical fluctuations associated with 
the formation of broken-symmetry state by explicitly introducing 
symmetry-breaking potentials or quantum measurements.

In the absence of the symmetry-breaking potential, 
the many-body ground state, obtained by the diagonalization 
of the symmetry-preserving Hamiltonian, is found to be fragile against 
a formation of a localized state. The localized state corresponds to 
a superposition state of quasidegenerate low-lying modes.  
The localization is also reflected in the enhancement 
in the two-body correlation.

In Sec.~\ref{MB_pot1}, we have studied the many-body localized state 
in the presence of an infinitesimal 
symmetry-breaking potential. We have shown that the localization of 
the many-body ground state begins when the energy 
scale of the symmetry-breaking potential becomes of the 
same order of the energy gap between 
the ground and the first excited state of the symmetry-preserving 
Hamiltonian, i.e., $1/N$. We have evaluated center-of-mass fluctuations 
in the regime of the crossover between 
the state with the translation symmetry and the broken-symmetry state, 
and found that these fluctuations are the origin of the depletion of the condensate fraction. 
In the presence of the double-well potential, 
we have shown that as the magnitude of the potential is 
increased, the ground state becomes a Schr\"{o}dinger's cat state, i.e., 
the macroscopic superposition of two localized states.

In a manner similar to the single-potential case, 
repeated quantum measurements also cause the crossover 
from the uniform state to the localized state with 
significant enhancement in the condensate fraction. 
We have shown that the center-of-mass and the angular-momentum fluctuations 
obey the uncertainty relation (\ref{mea_uncertainty}) during quantum measurements.

In the latter part of this paper, we have developed 
an analytic method to treat the many-body states with 
quantum fluctuations. 
Starting from the broken-symmetry ground state of the GP equation, 
we have taken into account the translation mode of the soliton as a Bogoliubov fluctuation 
in the presence of an infinitesimal symmetry-breaking potential. 
We have found that the angular-momentum fluctuations and the condensate fraction of 
the Bogoliubov ground state very well agrees with those of the numerically obtained 
many-body localized state. 

The effects of the quantum measurement of the bright soliton 
on the localization have been studied by developing a generalized mean-field theory, 
where we introduce a semiclassical distribution 
function of the center-of-mass position of the soliton. 
Assuming as an initial state an isotropic superposition of soliton states, 
we have argued that the measurement process 
reduces the center-of-mass fluctuation and 
selects a soliton state having a definite center of mass. 
The results obtained by the extended mean-field theory and those obtained by 
the numerical simulation using the symmetry-preserving Hamiltonian 
agree excellently in terms of 
the angular-momentum fluctuation, 
the center-of-mass fluctuation, 
and the condensate fraction. 

We have thus shown that 
the critical fluctuations associated with the formation of 
a localized BEC with an attractive interaction 
arise from quantum fluctuations in the position of 
the matter wave, which is also interpreted as the 
depletion of the condensate to the translation modes of the bright soliton.


\section{Acknowledgements}
This work was supported by Grant-in-Aids for Scientific Research (Grant
No. 15340129, No. 17071005, and No. 17740263) and by a 21st Century COE
program at Tokyo Tech ``Nanometer-Scale Quantum Physics,'' from the
Ministry of Education, Culture, Sports, Science and Technology of Japan. 
M.U. acknowledges support by a CREST program 
of the JST. 

\appendix
\section{Derivation of Operator Relation~(\ref{Op})}\label{adx}

In order to prove the relation (\ref{Op}), we disentangle two operators 
\bq
\hat{\cal O}_1&\equiv &e^{-\frac{r}{2}(\hat{T}^{\dagger 2}-\hat{T}^2)},\\
\hat{\cal O}_2&\equiv &e^{-\frac{\eta}{2}(\hat{T}^{\dagger}+\hat{T})^2}. 
\eq
We define an unnormalized state
\bq
|\alpha\rangle\equiv e^{\alpha\hat{T}^{\dagger}}|{\rm vac}\rangle, 
\eq
where $\alpha$ is a $c$ number, and 
define the expectation value of $\hat{\cal O}_j$ ($j=1,2$) with respect to 
the state $|\alpha\rangle$ as 
\bq
{\cal Q}_{j}\equiv \langle \alpha|\hat{\cal O}_j|\alpha\rangle. 
\eq
The derivatives of ${\cal Q}_j$ with respect to $\alpha$ and $\alpha^*$ are given by 
\bq
\frac{\partial{\cal Q}_j}{\partial\alpha}=\langle\alpha|\hat{\cal O}_j\hat{T}^{\dagger}|\alpha\rangle
=\langle\alpha|\hat{\cal O}_j\hat{T}^{\dagger}\hat{\cal O}_j^{-1}\hat{\cal O}_j|\alpha\rangle,\\
\frac{\partial{\cal Q}_j}{\partial\alpha^*}=\langle\alpha|\hat{T}\hat{\cal O}_j|\alpha\rangle
=\langle\alpha|\hat{\cal O}_j\hat{\cal O}^{-1}_j\hat{T}^{\dagger}\hat{\cal O}_j|\alpha\rangle.
\eq
By using the relations
\bq
\hat{\cal O}_1^{-1}\hat{T}\hat{\cal O}_1&=&\hat{T}\cosh r - \hat{T}^{\dagger} \sinh r,\\
\hat{\cal O}_1\hat{T}^{\dagger}\hat{\cal O}_1^{-1}&=&\hat{T}^{\dagger}\cosh r + \hat{T} \sinh r,\\
\hat{\cal O}_2^{-1}\hat{T}\hat{\cal O}_2&=&\hat{T}-\eta(\hat{T}+\hat{T}^{\dagger}),\\
\hat{\cal O}_2\hat{T}^{\dagger}\hat{\cal O}_2^{-1}&=&\hat{T}^{\dagger}-\eta(\hat{T}+\hat{T}^{\dagger}), 
\eq
we obtain
\bq
\frac{\partial{\cal Q}_1}{\partial\alpha}&=&
\alpha^*\cosh r{\cal Q}_1+\sinh r\frac{\partial{\cal Q}_1}{\partial\alpha^*},\label{Q1}\\
\frac{\partial{\cal Q}_1}{\partial\alpha^*}&=&
\alpha\cosh r{\cal Q}_1-\sinh r\frac{\partial{\cal Q}_1}{\partial\alpha},\label{Q1d}\\
\frac{\partial{\cal Q}_2}{\partial\alpha}&=&
(1-\eta)\alpha^* {\cal Q}_2-\eta\frac{\partial{\cal Q}_2}{\partial\alpha^*},\label{Q2}\\
\frac{\partial{\cal Q}_2}{\partial\alpha^*}&=&
(1-\eta)\alpha {\cal Q}_2-\eta\frac{\partial{\cal Q}_2}{\partial\alpha}\label{Q2d}. 
\eq
The differential equations for ${\cal Q}_j$ are then given by 
\bq
\frac{\partial{\cal Q}_1}{\partial\alpha}&=&
\frac{\alpha ^*+\alpha \sinh r}{\cosh r}{\cal Q}_1,\\
\frac{\partial{\cal Q}_1}{\partial\alpha^*}&=&
\frac{\alpha -{\alpha}^*\sinh r}{\cosh r}{\cal Q}_1,\\
\frac{\partial{\cal Q}_2}{\partial\alpha}&=&
\frac{1}{1+\eta}(\alpha^*-\eta\alpha){\cal Q}_2,\\
\frac{\partial{\cal Q}_2}{\partial\alpha^*}&=&
\frac{1}{1+\eta}(\alpha-\eta\alpha^*){\cal Q}_2.
\eq
By integrations of these equations, the expectation values ${\cal Q}_1$ and ${\cal Q}_2$ are found to be
\bq
{\cal Q}_1 &\propto& 
e^{-\frac{1}{2}\tanh r\alpha^{* 2}+\frac{1}{\cosh r}|\alpha|^2-\frac{1}{2}\tanh r\alpha^2},\nonumber\\
\ \\
{\cal Q}_2 &\propto&
e^{-\frac{1}{2}\frac{\eta}{1+\eta}\alpha^{* 2}+\frac{1}{1+\eta}|\alpha|^2-\frac{1}{2}\frac{\eta}{1+\eta}\alpha^2}.
\eq
The operators $\hat{\cal O}_j$ are thus written as 
\bq
\hat{\cal O}_1&\propto & 
e^{-\frac{1}{2}\tanh r\hat{T}^{\dagger 2}}
e^{-\ln (\cosh r) \hat{T}^{\dagger}\hat{T}}
e^{\frac{1}{2}\tanh r\hat{T}^2},\\
\hat{\cal O}_2&\propto & 
e^{-\frac{\eta}{2(1+\eta)}\hat{T}^{\dagger 2}}
e^{-\ln (1+\eta)\hat{T}^{\dagger}\hat{T}}
e^{-\frac{\eta}{2(1+\eta)}\hat{T}^2}, 
\eq
and we finally obtain 
\bq
e^{-\frac{r}{2}(\hat{T}^{\dagger 2}-\hat{T}^2)}|{\rm vac}\rangle\!\!&=&\!\!
{\cal N}_1 e^{-\frac{1}{2}\hat{T}^{\dagger 2}\tanh r}|{\rm vac}\rangle,\\
e^{-\frac{\eta}{2}(\hat{T}+\hat{T}^{\dagger})^2}|{\rm vac}\rangle\!\!&=&\!\!
{\cal N}_2 e^{-\frac{\eta}{2(1+\eta)}\hat{T}^{\dagger 2}}|{\rm vac}\rangle, 
\eq 
where the normalization constants ${\cal N}_j$ are determined by 
\bq
1&=&\langle {\rm vac}|\hat{\cal O}_1^{-1}\hat{\cal O}_1|{\rm vac} \rangle
={\cal N}_1^2\sum_{n}\frac{2n!}{(n!)^2}\left(-\frac{1}{2}\tanh r\right)^{2n}\nonumber\\
&=&{\cal N}_1^2\cosh r, \\
1&=&\langle {\rm vac}|\hat{\cal O}_2^{-1}\hat{\cal O}_2|{\rm vac} \rangle
={\cal N}_2^2\sum_{n}\frac{2n!}{(n!)^2}\left(-\frac{1}{2}\frac{\eta}{1+\eta}\right)^{2n}\nonumber\\
&=&{\cal N}_2^2(1+\eta). 
\eq

\section{Evaluation of Angular-Momentum Fluctuation in the Bogoliubov Ground State}\label{AM_fluct}

\begin{figure}
\includegraphics[scale=0.45]{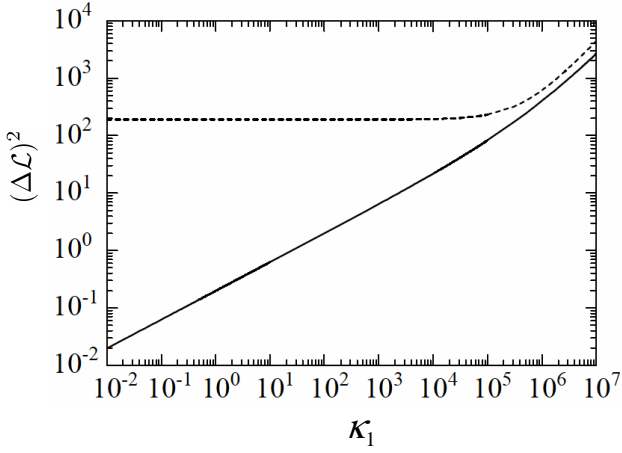}
\caption{
Width of the angular-momentum fluctuations in the GP ground state 
$|\Psi^{\rm (GP)}\rangle$ (dashed curve) 
and that in the state 
$e^{\frac{r}{2}(\hat{T}^{\dagger 2}-\hat{T}^2)}|{\rm vac}\rangle$ 
(solid curve). 
}
\label{fig_mf_bog}
\end{figure}

We justify the approximation, Eq.~(\ref{P_Beta}) with Eq.~(\ref{P_width_Beta}), 
that has been used in order to evaluate the angular-momentum fluctuation 
$(\Delta {\cal L}^{\rm (B)})^2$ in the Bogoliubov ground state 
\bq
|\Psi^{\rm (B)}\rangle=e^{\frac{r}{2}(\hat{T}^{\dagger 2}-\hat{T}^2)}
\frac{1}{\sqrt{N!}}(\alpha_0 \hat{c}_0^{\dagger}+\sqrt{2}\alpha_1 \hat{S}^{\dagger})^N|{\rm vac}\rangle. 
\eq
The angular-momentum distribution of the GP ground state 
$|\Psi^{\rm (GP)}_{\theta_{\rm c.m.}=0}\rangle=
\frac{1}{\sqrt{N!}}(\alpha_0\hat{c}_0^{\dagger}+\sqrt{2}\alpha_1\hat{S}^{\dagger})^N |{\rm vac}\rangle$ 
is calculated to give 
\bq\label{gp_am_fluct}
|\langle {\cal L}|\Psi^{\rm (GP)}_{\theta_{\rm c.m.}=0}\rangle|^2
&\propto& 
\sum_{n=1}^{N/2}\frac{1}{(N-{\cal L}-2n)!({\cal L}+n)!n!}\left(\frac{|\alpha_1|}{|\alpha_0|}\right)^{4n}\nonumber\\
&\simeq& \frac{1}{\sqrt{2\pi(\Delta{\cal L}^{\rm (GP)})^2}}e^{-\frac{{\cal L}^2}{2(\Delta {\cal L}^{\rm (GP)})^2}},
\eq
where $\alpha_0$ and $\alpha_1$ are given in Eqs.~(\ref{P_GPa1}) and (\ref{P_GPa0}), 
and the width is given by 
\bq\label{ApGP}
(\Delta {\cal L}^{\rm (GP)})^2=\frac{2N}{2+|\alpha_0|^2/|\alpha_1|^2}. 
\eq
On the other hand, from the form of Eq.~(\ref{P_Bgs}), the angular-momentum-fluctuation operator 
$e^{\frac{r}{2}(\hat{T}^{\dagger 2}-\hat{T}^2)}$ approximately gives the factor 
\bq\label{b_am_fluct}
\exp\left[-\frac{{\cal L}^2}{2(\Delta {\cal L}^{\rm (B)})^2}\right]
\eq
to the state with an angular momentum ${\cal L}$, 
where the width $(\Delta {\cal L^{\rm (B)}})$ is given in Eq.~(\ref{P_width_Beta}). 
The angular-momentum distribution of $|\Psi^{\rm (B)}\rangle$ is 
given by the product of Eqs.(\ref{gp_am_fluct}) and (\ref{b_am_fluct}). 
We compare Eqs.~(\ref{ApGP}) and (\ref{P_width_Beta}) in Fig.~\ref{fig_mf_bog}, where 
the former is much larger than the latter for $\kappa_1 \lesssim 10^4$. 
Therefore, the contribution from Eq.~(\ref{gp_am_fluct}) is negligible, 
and the angular-momentum distribution of $|\Psi^{\rm (B)}\rangle$ can be 
approximated by Eq.~(\ref{b_am_fluct}).



\begin{thebibliography}{100}
\bibitem{rotation}
When the system is subjected to a rotating drive, 
the degeneracy with respect to $E_{\pm {\cal L}}$ is lifted. 
For a strong coupling regime the angular momentum 
${\cal L}_{\rm g}$ in the ground state can increase. The superposition for the localized state 
in this case is given by 
$|\Psi_{\theta}^{(\varepsilon_1)}\rangle 
= e^{-i\hat{L}\theta} 
\left[\beta_0 |{\cal L_{\rm g}}\rangle 
+ \sum_{{\cal L}>0} \beta_{\cal L} 
(|{\cal L_{\rm g}}+{\cal L}\rangle
 +|{\cal L_{\rm g}}-{\cal L}\rangle)\right]$ instead of Eq.~(\ref{MBLoc}). 
The angular momentum thus fluctuates around ${\cal L}_{\rm g}$, 
and the center of mass also fluctuates with the width 
$\Delta\theta_{\rm c.m.} \propto (\Delta{\cal L})^{-1}$.
\end{thebibliography}
\end{document}